\documentclass[a4paper,amsmath,amssymb]{jpconf}
\usepackage{graphicx}
\begin{document}

\newcommand{\Ns}{{N_s}}
\def\NOTES#1 {{}}      %%% for permanent supplementary annotations
\def\OMIT#1 {{}}   
\def\remark#1 {{\sl #1}}   
\def \eqr#1{(\ref{#1})}
\def \be {\begin{equation}}
\def \eeq {\end{equation}}
\def \bea {\begin{eqnarray}}
\def \eea {\end{eqnarray}}
\def \la {{\langle}}
\def \ra {{\rangle}}
\def \nn {\hat{\bf n}}
\def \hh {{\bf h}}
\def \ss {{\bf s}}
\def \HH {{\cal H}}
\def \QQ {{\bf Q}}
\newcommand{\Jcal}{{\mathcal{J}}}
\newcommand{\tlambda}{{\tilde \lambda}}
\newcommand{\sqrtthree}{``$\sqrt{3}\times\sqrt{3}$''}
\newcommand{\Heff}{\HH_{\rm eff}}
\newcommand{\xlink}{x_{\rm link}}
\newcommand{\xkag}{x_{\rm kag}}
\def \PP {{\bf P}}
\def \DD {{\bf D}}
\def \EE {{\bf E}}
\def \rr {{\bf r}}
\newcommand{\alphaD}{\alpha^{\rm D}}
\newcommand{\alphaE}{\alpha^{\rm E}}
\newcommand{\alphaDE}{\alpha^{\rm DE}}
\newcommand{\sigmaDE}{\sigma^{\rm DE}}
\newcommand{\sigmaE}{\sigma^{\rm E}}

\title{Depleted pyrochlore antiferromagnets}

\author{Christopher L. Henley}

\address{Dept. of Physics, Cornell Univ., Ithaca NY 14853-2501 USA}

\ead{clh@ccmr.cornell.edu}

\begin{abstract}
I consider the class of ``depleted pyrochlore'' lattices
of corner-sharing triangles, made by removing spins from a
pyrochlore lattice such that every tetrahedron loses exactly one.
Previously known examples are the ``hyperkagome'' and ``kagome staircase''.
I give criteria in terms of loops for whether a given depleted lattice
can order analogous to the kagome \sqrtthree~ state, 
and also show how the pseudo-dipolar correlations (due to local
constraints) generalize to even the random depleted case.
%%% abstract at most 200 words. 
\end{abstract}

Consider ``bisimplex'' antiferromagnets~\cite{Hen01}, meaning that {\it every} spin is 
shared between exactly two triangles or tetrahedra\OMIT{ (or even just a bond)}.
Let there be only the nearest-neighbour coupling, so the system is highly frustrated.
%%%%%%%%%%%%%%%%%%%%%%%%%%%%%%%%%%%%
\OMIT{(meaning a large (near) degeneracy of states).}
%%%%%%%%%%%%%%%%%%%%%%%%%%%%%%%%%%%%
After the kagom\'e~\cite{chalker-kagome}, pyrochlore~\cite{An56,moessner98}, 
and garnet~\cite{petrenko} lattices, further such systems were
discovered such as the ``kagom\'e staircase'' 
(realised in e.g.  Ni$_3$V$_2$O$_8$~\cite{kag-staircase})
and the ``hyperkagom\'e'' lattice~\cite{hopkinson,lawler}
(realised in Na$_4$Ir$_3$O$_8$~\cite{hyperkag}).
Both magnetic lattices are obtained from
the pyrochlore lattice by removing 1/4 of the sites so as to leave a network
of corner-sharing triangles.  This paper considers the entire family of such 
``depleted'' structures, and the equilibrium states of classical or semiclassical 
spins on them.~\footnote{
%%%%%%%%%%%%%%%%%%
I will not consider other depleted lattices, except to observe the
hidden symmetry (considering nearest-neighbour bonds) whereby 
the hyperkagom\'e lattice~\cite{hyperkag} is equivalent to a 
garnet lattice~\cite{petrenko}, is reminiscent of the
hidden equivalence of the 1/5-depleted square lattice of 
CaV$_4$O$_9$~\cite{cav4o9} to the 4-8 lattice~\cite{4-8}.
}

The pyrochlore lattice is most simply visualised as the ``medial graph''
(bond midpoints) of a diamond lattice.  
%%%%%%%%%%%%%%%%%%%
\OMIT{(In general, that is the ``parent lattice''; the kagom\'e lattice
is the medial graph of the honeycomb lattice as parent lattice.)}
%%%%%%%%%%%%%%%%%%%
A depleted lattice is made by placing a dimer covering on the diamond lattice bonds, 
then removing spins from the covered sites.
%%%%%%
\OMIT{That is equivalent to diluting the pyrochlore lattice, such that 
exactly one spin is taken from each tetrahedron.}
%%%%%%
This constraint is plausible: in real spinels 
with magnetic B sites (= a ``pyrochlore'' magnetic lattice),  
dilution is achieved by substituting a nonmagnetic species X on B sites.
Due to size (or maybe charge) imbalance, X ions repel; hence the lattice
gas of X ions maps to an Ising antiferromagnet~\cite{An56}
in a field. That is itself a highly frustrated problem, with a degenerate ground state
(all those dimer coverings) -- provided the lattice gas has only
nearest neighbour interactions. Further neighbour ion terms presumably 
select a specific depletion pattern.

\OMIT{
Our interest in depleted lattices stems from the following
facts.  (i) They are natural ways that an oxide structure
can be modified so as to satisfy valence requirements, 
making substitutions by a different cation, or a charge order 
in which the same element has two different possible ionization 
states (one of them being nonmagnetic).  
(ii)  Randomly depleted antiferromagnets with nearest-neighbor interactions 
retain most of the interesting properties of uniformly frustrated lattices.
(iii) Not infrequently,
there are depletion patterns such that -- if only first neighbor
bonds are taken into account -- the depleted lattice
has a hidden equivalence to some known,  more symmetric structure.}

After some examples of depleted lattices, I address two questions 
(for the periodic {\it and} random cases):
(i) What is the pattern of magnetic order  (if any);
(iii) How do we generalize the disordered classical liquid with pseudo-dipolar
correlations due to the constraints? 

\section{Regularly depleted pyrochlore lattices}
\label{sec:examples}

In this section, I catalog some highly symmetrical dimer
coverings of the diamond lattice, each of which specifies 
a different depleted pyrochlore lattice.  

These can be conveniently 
be visualized in two possible ways: (i)  projecting the conventional 
cubic cell (containing four diamond sites, i.e. four pyrochlore tetrahedra) 
in the (100) direction; or, 
(ii) expressing the diamond lattice as a stacking of puckered 
honeycomb layers, in which the (odd) sites have an additional bond 
extending upwards (downwards) to the next layer.

\OMIT{I should rank the packings according to how uniform
the dimer centers are spaced, since that is presumably favored
e.g. by Coulomb interactions. Of course, since every diamond vertex has
exactly one dimer endpoint, {\it all} dimer coverings are rather
uniform.}

\subsection{Cubic conventional cell}

Consider the family of patterns preserving the periodicity of the
conventional cubic cell, with four dimers per cell.
\OMIT{Thus we have to place four dimers in one
cell, with periodic boundary conditions.}
There are only three symmetry-inequivalent ways to place them.  
Pattern 1 has all dimers oriented the same; the diamond lattice
separates into disconnected (puckered) honeycomb layers,  and the
spins form stacks of uncoupled kagom\'e lattices, 
as induced by a field in ``kagom\'e ice''~\cite{kag-ice}.
%%%%%%%%%%%
\OMIT{(In that case, the lattice is depleted not by removing spins, but by 
freezing out the spins whose easy axis is directly aligned with a field)}

Pattern 2 has two dimers in one orientation
and two in another orientation. 
This again separates the lattice into disjoint slabs,
now transverse to a (110) axis.
\OMIT{There should be a figure showing the slab end-on.}
The depleted lattice is a ``kagom\'e staircase'',
which has the topology of a
kagom\'e lattice, but folded so the hexagons alternate
between two (111) type orientations. 
%%% differing by $71^\circ=\cos^{-1}(1/3)$.
%%%%%%%%%%%%%%%%%%%%%%%%%%%%%%%%%%%%%%%%%%%%%
Imagine a canonical spinel AB$_2$O$_4$  with cubic lattice constant 
$a_c\approx 8.2$~\AA; depletion by B-site vacancies gives
the formula A$_1$B$_{1.5}$O$_4$ with $x=1/4$.
Pattern 2 makes the structure orthorhombic, with
$a\approx a_c/sqrt{2}$, $b\approx \sqrt{2}a_c$, and $c\approx a_c$.
This nearly describes the ``kagom\'e staircase'' compounds
e.g. VCo$_{1.5}$O$_4$, except each successive slab is slid 
a half cell relative to the one before. 

Pattern 3 uses each of the four possible orientations once, 
yielding the ``hyperkagom\'e'' arrangement with cubic symmetry.
%%%%%%%%%%%%%%%%%
\OMIT{The dimers themselves form a ``trillium'' lattice of 
equilateral triangles with three corners shared at each vertex~\cite{trillium}.
The bond network is topologically
equivalent to the Laves graph of degree three (better known as the
gyroid graph) and so the spin lattice
is equivalent to the (more symmetric) garnet lattice, as mentioned 
in Ref.~\cite{hyperkag}.
Note the garnet lattice was previously called 
hyperkagom\'e~\cite{old-hyperkag}.}
%%%%%%%%%%%%%%%%%%
as realized~\cite{hyperkag} in the spinel Na$_4$Ir$_3$O$_8$, i.e.
``(Na$_{1.5}$)$_1$(Ir$_{3/4}$Na$_{1/4}$)$_2$O$_4$''
in our framework, \OMIT{where does the extra 0.5 Na fit?}. 

\OMIT{The cubic cases are conveniently visualized by projecting along
a (100) axis.}

\subsection{Kagom\'e layer stacking approach}

\begin{figure}[h]
 \begin{minipage}{7pc}
 \includegraphics[width=7pc]{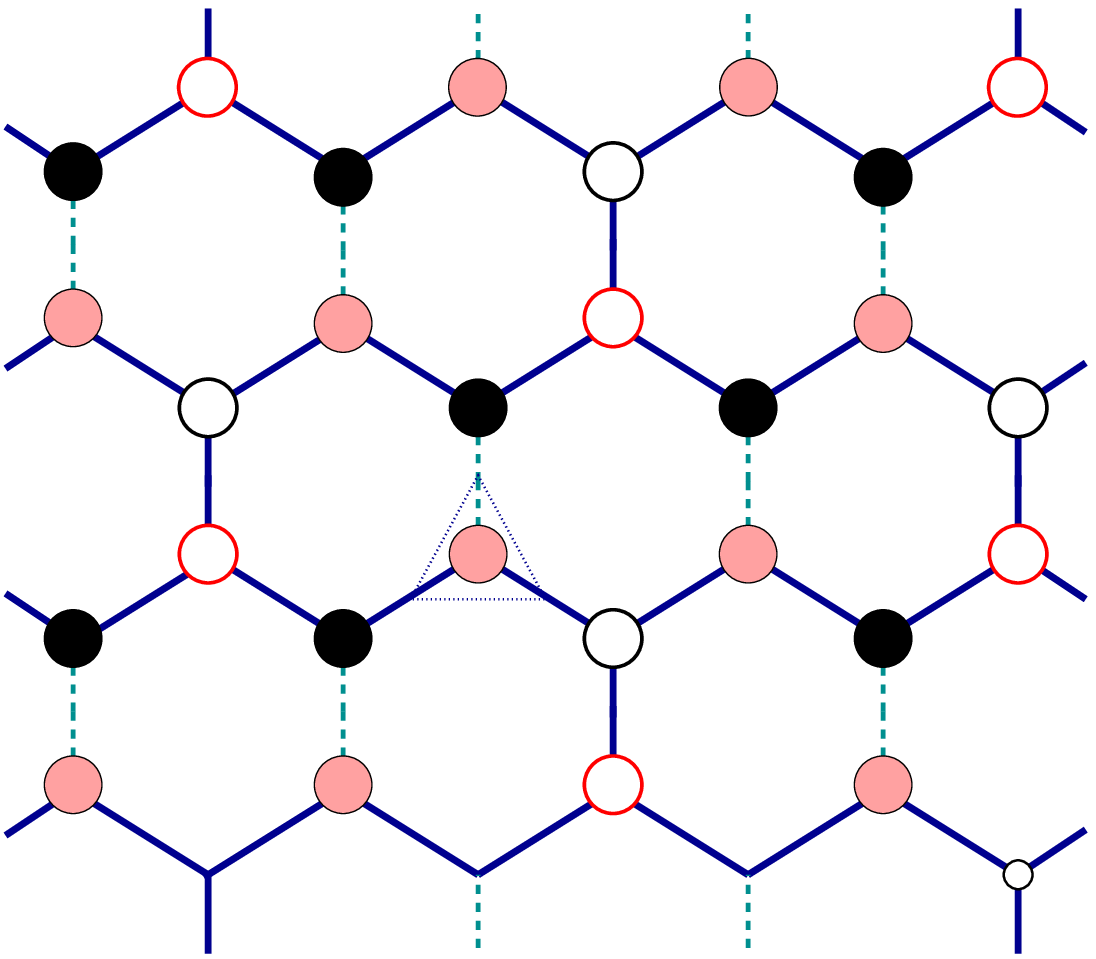}
 \end{minipage}\hspace{2pc}%
 \begin{minipage}{7pc}
 \includegraphics[width=7pc]{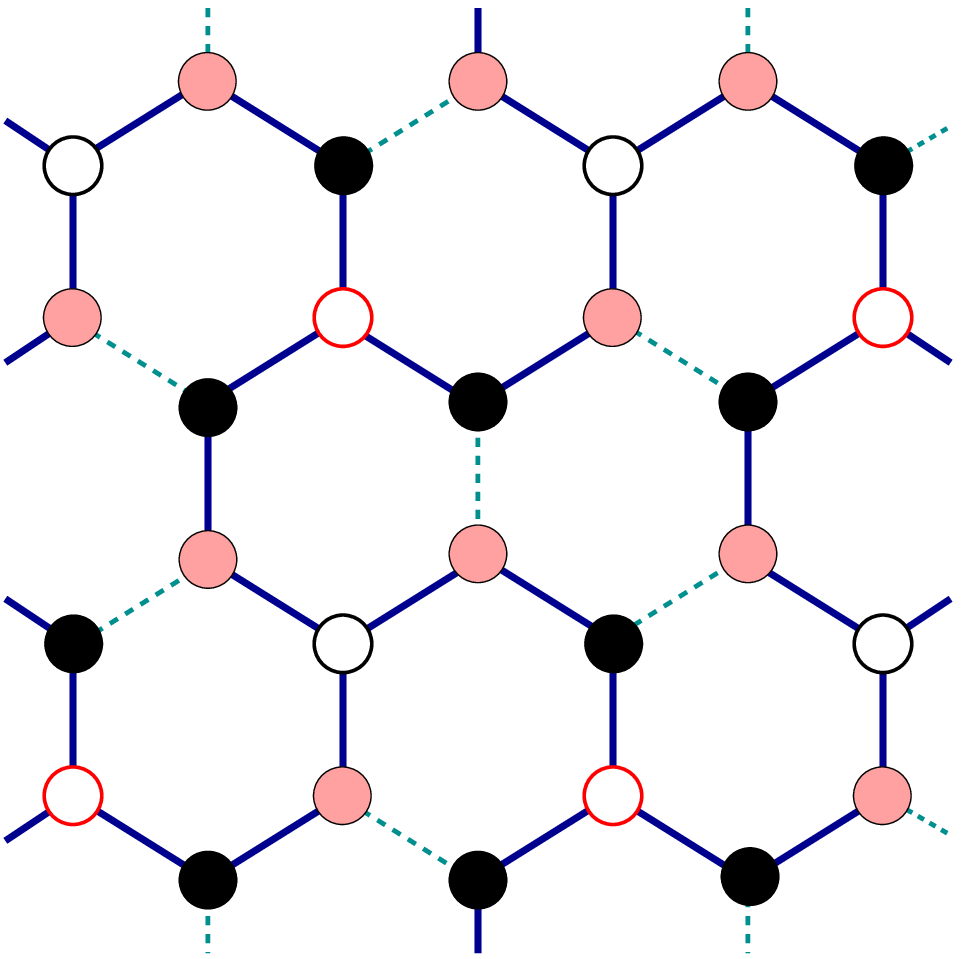}
 \end{minipage}\hspace{2pc}%
 \begin{minipage}{7pc}
 \includegraphics[width=7pc]{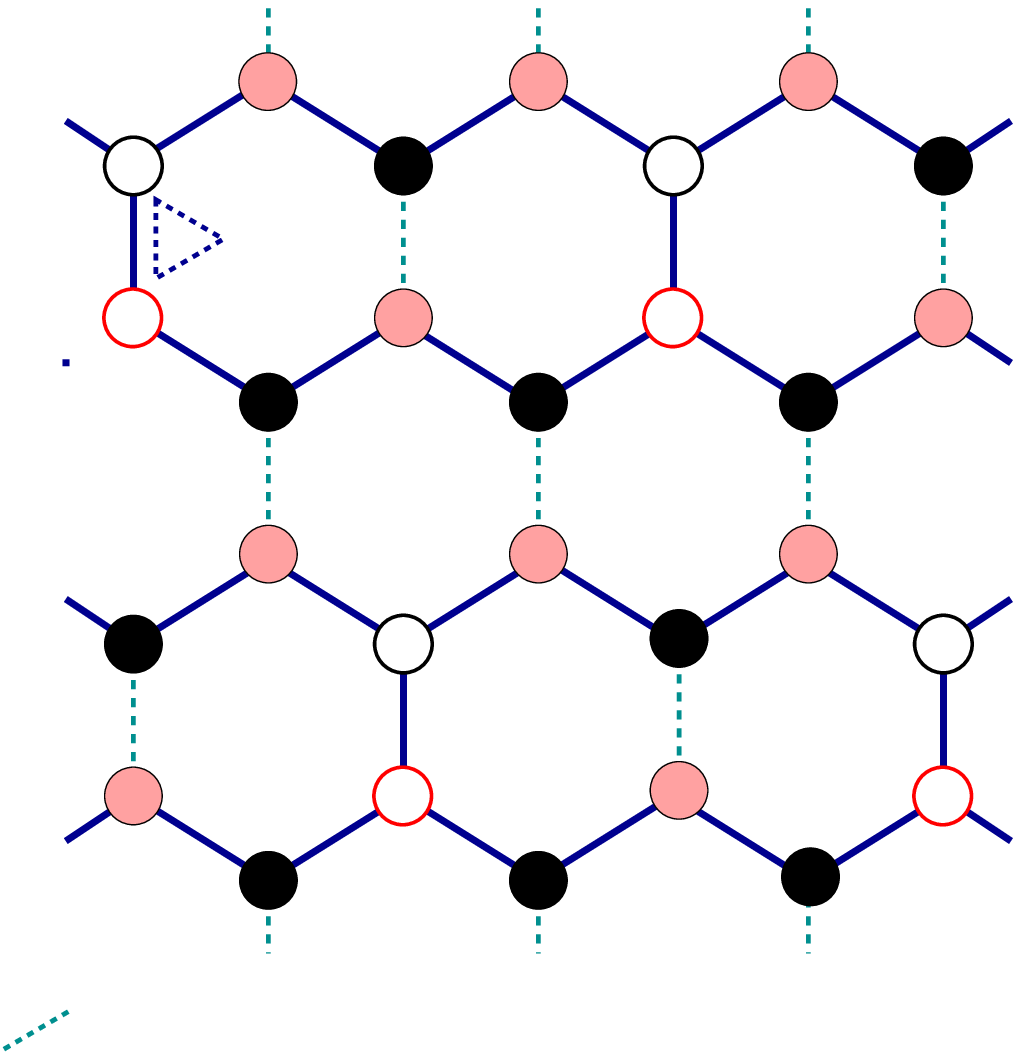}
 \end{minipage}\hspace{2pc}%
 \begin{minipage}{7pc}
 \includegraphics[width=7pc]{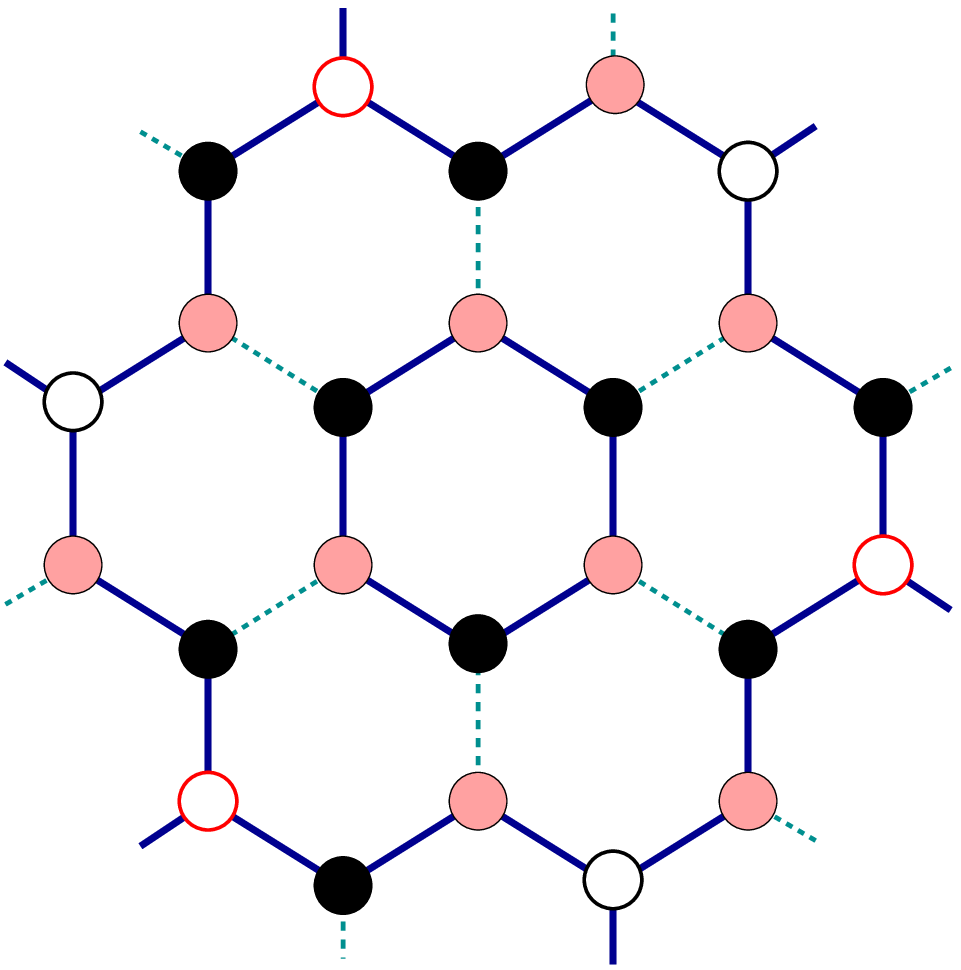}
 \end{minipage}
 \caption{\label{fig:lattices} Depleted lattices built by stacking
puckered layers along a 3-fold axis. 
Diamond lattice are shown bonds as lines (in the layer) or as circles (for linking
bonds: up=shaded,  down=black).
Removed bonds are indicated by dashed lines or empty circles.
In (a) and (c), successive layers are
stacked with a $2\pi/3$ rotation around the dotted triangle.
(a). a $\sqrt 3 \times \sqrt 3$ pattern; 
(b). hyperkagom\'e lattice;
(c). another $2 \times 2$ pattern;
(d). a $\sqrt 7 \times \sqrt 7$ pattern.
\OMIT{Not included: lattice of non-orientable 6-loops (marked by arrows).}
}
 \end{figure}

I will focus on the layered patterns made by viewing the
diamond lattice as a stack of puckered honeycomb layers 
(a kagom\'e layer of spin sites)
connected by vertical linking bonds (forming a triangular lattice).
Say a fraction $\xlink$ of linking bonds is depleted;
then the honeycomb layer has depletion $\xkag\equiv (1-\xlink)/3$;
its depletion pattern is a dimer covering 
having monomers at the endpoints of the depleted linking bonds.
If either kind of layer lacks threefold point symmetry, we restore it by 
stacking the layers with a rotation (producing a screw axis).
%%%%%%%%%%%%%%%
\OMIT{The examples selected to present here in Fig.~\ref{fig:lattices} and 
Table~\ref{tab:lattices} place dimers on linking bonds in a triangular superlattice.}
%%%%%%%%%%%%%%%

\OMIT{Let $\xlink=1$; this produces disconnected kagom\'e layers,
the same as cubic cell pattern 1.
Of course, pattern 1 may also be generated with $\xlink=0$, $\xkag=1/3$,
when all the honeycomb-layer dimers are parallel.
With $\xlink=0$, $\xkag=1/3$, we could instead use a 3-fold symmetric pattern 
in each honeycomb layer, stacked in various fashions.
We can get $\xlink=1/2$ by dimers in alternate rows
of the triangular lattice of the linking layer;  the pattern of the dimers 
in honeycomb layers $(\xkag=1/6$) is forced. To restore symmetry, we rotate
the linking layers by $2\pi/3$ in successive layers. }

\begin{center}
\begin{table}[h]
\centering
\caption{\label{tab:lattices} Some depleted lattices, with 
properties of the shortest loops
(Loop density is per site; in the tags, interlayer bonds are
underlined for comparison to Fig.~\ref{fig:lattices}.)}
\begin{tabular}{lllrlll}
\br
Fig.~\ref{fig:lattices}  & $\xlink$ & $\xkag$  &  $L_{\rm loop}$ & loop    & 
      Colouring & Loop is \\
part   &   &   &  & density &  tags  &  colourable? \\
\mr
a)     &  1/3   &  2/9     &    8  & 1/2   &   $bbbs\underline{b}bb\underline{b}$ & no\\
b)     &  1/4   &  1/4     &   10  & 3/4   &   $(bbsb\underline{s})^2$  &  yes \\
c)     &  1/4   &  1/4     &   8  & 1/16   &   $bb\underline{b}bb\underline{s}sb$ & no \\
d)     &  1/7   &  2/7     &   6  & 1/14   &   $bbbbbb$ &  yes \\
\br
\end{tabular}
\end{table}
\end{center}

 %%% say "angle=90" to rotate it

% \begin{figure}[h]
% \begin{minipage}{10pc}
% \includegraphics[width=10pc]{tri3.eps}
% \end{minipage}\hspace{2pc}%
% \begin{minipage}{10pc}
% \includegraphics[width=10pc,angle=90]{tri7B.eps}
% \end{minipage}\hspace{2pc}%
% \begin{minipage}{10pc}
% \includegraphics[width=10pc]{unori.eps}
% \end{minipage}
% \caption{\label{fig:lattices} Depleted lattices constructed by stacking
%layers.  Figure shows bonds of the diamond lattice; lines are bonds in the
%layer,  and circles are bonds up (shaded) or down (black).
%Removed bonds are indicated by dashed lines or empty circles.
%(a). $\sqrt 3 \times \sqrt 3$ pattern; successive layers are
%stacked with a $2\pi/3$ rotation around the dotted triangle.
%The out-of-plane loop is orientable, but not properly colourable.
%(b). $\sqrt 7 \times \sqrt 7$ pattern (c). Orthorhombic cell
%containing non-orientable 6-loop (marked by arrows).}
% \end{figure}

%% \begin{figure}[h]
%% \begin{minipage}{14pc}
%% \includegraphics[width=14pc]{tri3.eps}
%% \caption{\label{label}Figure caption for first of two sided figures.}
%% \end{minipage}\hspace{2pc}%
%% \begin{minipage}{14pc}
%% \includegraphics[width=14pc]{tri7v2.eps}
%% \caption{\label{label}Figure caption for second of two sided figures.}
%% \end{minipage}
%% \end{figure}

Fig.~\ref{fig:lattices} shows some examples; there are many more,
e.g. in Fig.~\ref{fig:lattices}(c) a different honeycomb dimer pattern
could be used.  Fig.~\ref{fig:lattices}(b) is just the 
cubic ``hyperkagom\'e'' structure. 
%%%%%%
With $\xlink=0$ and $\xkag=1/3$ we get a dimer covering
of the honeycomb lattice. 
\OMIT{ These are some of the most  interesting cases.  
Within a honeycomb layer, 
we may either have a $\sqrt 3 \times \sqrt 3$ packing
or all dimers parallel. 
When we go from one layer to the next, we'll spiral 
in order to create 3-fold symmetry.
If we translated the parallel pattern (instead of spiraling)
in going to the next layer, we'd get
cubic-cell pattern 1. 
If we have the parallel arrangement in a layer and rotate it,
we get nice vertical spirals of dimers;  there are
lines of spins going in 3 directions, woven together, with
linking spins.  This is particularly symmetrical, in having only
two kinds of site (spin lines and linking spins).}

\OMIT{{\sl TO CHECK: In the $\sqrt 3 \times \sqrt 3$ arrangement, we have 
isolated hexagons within the layers; only the linking spins
can make an extended.  If we translate the hexagons in the
same direction from each layer to the next, we make a structure
aligned in slabs.  Are they (110) slabs?  Anyhow, it is not
the kagom\'e staircase.}}

\section{Magnetic ground state}
\label{sec:ground}

The magnetic Hamiltonian is assumed to be $\HH = J \sum_{\la ij\ra} \ss_i\cdot \ss_j$
where only nearest-neighbour isotropic interactions are included, and $\{ \ss_i \}$
are either classical unit vectors 
%% \OMIT (at a small or zero temperature $T$), 
or quantum spins with $S\gg 1$.  Let us review the known story 
for the kagom\'e or garnet/hyperkagom\'e) antiferromagnets.
The classical ground states are the (many) configurations in which the
spins differ by angles $2\pi/3$ on every triangle.  
It then transpires that (i) a (still highly degenerate) subset of ``coplanar'' 
states gets selected at harmonic order, which amounts to a 3-colouring; 
(ii) a specific coplanar ordered state is selected by anharmonic fluctuations, 
which was the \sqrtthree~ state
in the kagom\'e case~\cite{huse-rut,chubukov,chan}
and Lawler's state in the hyperkagom\'e/garnet case~\cite{lawler}.
Do these results generalise?
%%%
\OMIT{Specifically: we must check 
(i) are coplanar states possible? (answer: yes -- many!)
(ii) can the (generalised) recipe for an ordered state be satisfied
(answer: often, but not necessarily).}

\subsection{Coplanar states}

At harmonic order, thermal fluctuations 
(in the classical case~\cite{chalker-kagome,moessner98})
or quantum fluctuations~\cite{chubukov,ritchey} select 
``coplanar'' ground states, such that
spins in different triangles lie in the same plane of spin space.~\footnote{
%%%%%%%%%%%
The constraint counting argument of \cite{moessner98} 
carries over independent of how the triangles are arranged.
\OMIT{They showed that the relative selection 
diverges as $T\to 0$, hence one expects long-range order of this plane
in spin space.})
%%%%%%%%%%%%
}
This selection should carry  over to arbitrary corner-sharing
triangle networks. 
%%%%%%%%%%%%%
\OMIT{even the ``triangular (Husimi) cactus'' (see below)}
\OMIT{Crude approximations of the harmonic quantum correction,  along 
the lines of \cite{long89,hen-pyroc05,doron-bergman}, give an effective 
Hamiltonian of the form $(1/S) \sum |\ss_i \cdot \ss_j \times \ss_k |^2$.
I cut this because I expect the true answer is linear in the deviation angle.}
%%%%%%%%%%%%%%%%%%%%%

A coplanar ground state has every spin in one of three directions
so it is effectively a 3-colouring of the sites 
(equivalently, diamond-lattice bonds) by colours $A,B,C$, such that 
every triangle (i.e. bonds meeting at one diamond vertex) has one of
each colour.  
By K\"onig's theorem~\cite{graph-theory} of graph theory,
a 3-colouring exists on any bipartite graph
with coordination $z=3$, and hence on {\it any} depleted lattice.
%%%%%%%%%%%%%%%
\OMIT{(The depleted lattice is bipartite since the diamond lattice is.)
To be exact, K\"onig's theorem states that any bipartite 
graph of maximum degree (= coordination number) $z_{\rm max}$ can be
edge-coloured using only $z_{\rm max}$ colors. 
The bipartiteness is crucial: certain degree-3 graphs, known as
``snarks'' and commonly containing loops of length 5, cannot be
3-coloured.}
%%%%%%%%%%%%%%%
\OMIT{This also confirms that a $2\pi/3$ coplanar state is
always possible.}

Any 3-colouring of a depleted lattice can usefully be reimagined as a
4-colouring of the original pyrochlore lattice, with the fourth color
corresponding to the depleted sites.  Thus, any 3-colouring
in fact generates {\it four} different ways to build a depleted lattice
(along with a sample 3-colouring of each), depending on which color we
select for removal.  (In Lawler's state~\cite{lawler}, all four colours
are equivalent; each colour forms a ``trillium'' lattice~\cite{trillium}.)
%%%%%%%%%%%%%%%%%%%%%
\NOTES{More on Lawler's state. If the four colours represented four 
tetrahedrally related directions
in spin space, then all even tetrahedra have one handedness of the
spin pattern and the odd tetrahedra have the opposite handedness.}
%%%%%%%%%%%%%%%%%%%%%%%%

\subsection{Ising mapping and effective Hamiltonian}
\label{sec:Heff}

As is well known, at harmonic order all the coplanar states have
equivalent Hamiltonians~\cite{chalker-kagome,chubukov}. 
Hence, fluctuations distinguish among them only at {\it anharmonic} order.  
Observe too that coplanar configurations cannot distinguished on the 
loop-free ``cactus lattice'' -- the medial graph of a $z=3$ Bethe lattice -- 
since they are all symmetry-equivalent by permutations of the sites.  
{\it Loops} are essential to state selection~\cite{uzi,bergman}. 
%%%%%%%%%%%%%%%%%%%%%%%%%%%%%
\OMIT{That is why loop statistics are used to 
classify the zoo of depleted lattices (Table ~\ref{tab:lattices}.)}

On the kagom\'e lattice, coplanar states are more transparently
represented by ``chiralities'' $\eta_\alpha=\pm 1$, Ising variables
defined on triangle centers $\alpha$, and equal to $+1$ ($-1$);
if the colours $ABC$ run counterclockwise (clockwise) around the
triangle.  The colouring can be uniquely reconstructured (modulo
trivial symmetries) from $\{ \eta_\alpha \}$, but not every
configuration of $\{ \eta _\alpha \}$ corresponds to a colouring.
With approximations, one can obtain an effective
Hamiltonian of form 
$\Heff = - \sum _{\alpha\beta} \Jcal_{\alpha\beta} \eta_\alpha \eta_\beta$
   %%% \label{eq:Heff}
in the quantum case~\cite{chan}, and also  in the
classical case~\cite{hen-unpub} at small $T$, based on \cite{soft-Heff}.
The nearest neighbor Ising coupling $\Jcal_1 <0$, so the optimum state is 
an antiferromagnetic pattern of $\eta_\alpha$ on the honeycomb vertices.

What about $d=3$? Let the index $\alpha$
label triangles, or equivalently diamond-lattice vertices.
A gauge choice is necessary on every triangle 
to define which sense of its normal vector is ``up'',
before the spin chirality can be defined.
\OMIT{A natural gauge choice (if it works) is such that the axes of neighbouring
triangles have positive projection on each other.}
I conjecture that, in general, the sign of $\Jcal_1$
is the opposite of the projection of the normal vectors of the
respective triangles.  

Then the ground state has an alternating chirality 
pattern (which is always possible, since the diamond vertices are bipartite).
For the hyperkagom\'e lattice -- the most regular depleted example --
this gives the correct answer: Lawler's state, favored
in a large-$n$ calculation~\cite{lawler} and found in simulations~\cite{hopkinson}.)
Whereas the \sqrtthree~ state of the kagom\'e had 
alternating colours (e.g. $ABABAB$) around loops and triple colors 
($ABCABC...$) along lines, and has a nonzero ordering vector, Lawler's state 
has alternating colours along lines and ordering vector $Q=0$.  
\footnote{
%%%%%%%%%%%%%%%%%%%%%
Lawler's state {\it lacks} ``weathervane'' modes which were
sometimes considered as the key feature on the kagom\'e.}
%%% selecting the \sqrtthree~ state 
%%%%%%%%%%%%%%%%%%%%%
\OMIT{But the proper analog of the kagom\'e $Q=0$ state is presumably the hyperkagom\'e
state (at $Q\neq 0$) which, like it,  has a ferromagnetic pattern of chiralities.}

The basis for believing $\Jcal_1 <0$ is general is that
in the kagom\'e case, it was expressed~\cite{chan,hen-unpub} 
in terms of expectations of spin-wave fluctuations of ``soft'' modes,
which have only anharmonic-order restoring forces.
Such soft modes are generic to the coplanar state on any of our lattices.
They sum to zero on every triangle, a ``zero-divergence'' constraint that
implies generic power-law correlations of the fluctuations, whether 
classical~\cite{huse-pseudodip,clh-pyrice,isakov} or quantum~\cite{chan,hermele}.
We obtain $\Jcal_1 <0$ if the sign depends on orientation the same way it does asympotically.  
A caveat is that at root, the crucial spin-wave correlations must depend on
the loops (as noted above, only loops distinguish among coplanar states);
the pseudo-dipolar correlations are just a coarse-grained way to incorporate
the net effects of many long loops.  An alternate local derivation of $\Heff$, based
on a loop expansion~\cite{uzi,bergman,uzi-loop} might
better capture the differences among depleted lattices.
%%%%%%%%%%%%%%%%%%
\OMIT{(Some of those are poor in short loops.)}

\subsection{Colourability}

Let's call the lattice ``colourable'' if there is a 3-colouring satisfying 
the above rule.  That is true if and only if every fundamental loop
of diamond-lattice bonds is colourable.  Tag the each bond of the loop 
as ``$s$'' or ``$b$'' depending whether the corresponding site is
analogous to a ``straight'' or ``bent'' point in a path 
on the kagom\'e lattice.~\footnote{
%%%%%%%%%%%%%%%%%%%
Note that given step $i$ there
is always (at least) one non-depleted direction available at both steps
$i+1$ and $i-1$; if the path enters on one of these but does not leave
on the other, it is ``bent'', otherwise it is ``straight''. }
%%%%%%%%%%%%%%%%%%%
Surrounding a $b$ step the colouring alternates (e.g. $ABA$), as in a kagom\'e
hexagon, whereas surrounding an $s$ step it cycles (e.g. $ABC$), as on a straight
line of kagom\'e.  Let $\{j_1, j_2, ...\}$ be the steps in the loop tagged $s$;
it can be shown the colours  are consistent around the loop if and only if 
$\sum_m (-1)^{m+j_m} \equiv 0 $(mod 6).~\footnote{
%%%%%%%%%%%%%%%
Or you may just start colouring
(e.g.  $ABAC...$) in accordance with the rules mentioned, and see whether the loop
ends on the same colour it started with.}
%%%%%%%%%%%%%%%%%%%
\OMIT{The loop is unorientable if and only if the number of $s$ bonds is odd.}

\NOTES{{\sl Orientability --}
I will call a loop ``orientable'' 
when the ``neighbour'' gauge is well-defined;
that is, if the (gauge-invariant) product of the
projection of adjacent axes is positive.
If all fundamental loops are orientable, the effective Hamiltonian has a unique optimum
chirality configuration $\{ \eta_\alpha \}$, but that doesn't guarantee colourability.}

Colourability does not depend on just the loop shape, but also on the
orientations of the depleted dimers touching each site of the loop.
Consider 6-loops (the shortest that we may have) in one layer 
(like those shown in Fig.~\ref{fig:lattices}):
each vertex of the hexagon can be tagged by 0 (if the depleted dimer runs vertical
from it) or 1 (if the depleted dimer sticks outwards from the hexagon center).
%%%%%%%%%%%%%%%
\OMIT{A successive ``11'' pair implies an $s$ bond, all other pairs imply
$b$ bonds.}
%%%%%%%%%%%%%%%
Of the $2^6$ possible 6-loops, 45/64 are not colourable:
the seven classes $(111000)$, $(111010)$, $(110110)$, $(111110)$, 
$(110000)$, $(110100)$, and $(111100)$. 
%%%%%%%%%%%%%%%
\OMIT{Of these, the last three -- representing 24/64 of the 6-loops --
are not even orientable.  The non-orientable loops
have low point symmetry (also they frustrate global patterns which 
could incorporate their point symmetry), so the high-symmetry 
periodic depleted lattices are I found orientable.)}
%%%%%%%%%%%%%%%
The six colourable classes account for 19/64 of the 6-loops.

\section{Classical cooperative paramagnet}
\label{sec:coop-para}

I first review known properties of random ensembles with divergence-like constraints.
Say some bonds of a bipartite $z$-coordinated lattice are coloured (with the
same colour) 
such that $z'$ bonds are coloured at every vertex.
Construct a divergence-free vector field  by letting every bond 
have unit flux from the even to the odd endpoint if coloured, 
or flux $-z'/(z-z')$ if uncoloured; the local volume average
defines a coarse-grained ``polarization'' 
$\PP(\rr)$ with $\nabla\cdot \PP =0$
\cite{huse-pseudodip,clh-pyrice,isakov}.
In a random  ensemble of such colourings, the entropy density behaves as 
   \be 
          \sigma(\PP) = \sigma_0  - \frac{1}{2} \alpha |\PP|^2 + \hbox{\it higher order}
   \label{eq:entropy} 
   \eeq
Combined with the divergence constraint, \eqr{eq:entropy} implies
long-range, pseudo-dipolar correlations:
   \be
         \la  P_a (0) P_b(\rr) \ra \propto 
       \frac{d (r_a r_b/|\rr|^2) - \delta_{ab}} {\alpha |\rr|^d}
   \label{eq:pseudodip}
   \eeq
in $d$ dimensions.
Correlations of the physical degrees of freedom are generally proportional to 
\eqr{eq:pseudodip}.  Dimer models~\cite{huse-pseudodip}
are the case $z'=1$; ice models are the case $z=4$, $z'=2$.
A $z$-colouring produces $z-1$ flavours of $\PP(r)$ (one for each colour, minus
one linear dependency) hence $\PP(\rr)$ in that case is a $d \times (z-1)$ tensor.
The ground states of classical spins on a bisimplex lattice -- our original problem
-- satisfy the constraint $\sum _{i\in \alpha} \ss_i =0$~\cite{Hen01}; the polarization
in that case is a $d \times m$ tensor where $m$ is the number of spin
components, and here too we expect \eqr{eq:pseudodip}  describes the 
correlations~\cite{hopkinson,clh-pyrice,isakov}.

The problem at hand is a depleted pyrochlore lattice ($z=3$) -- periodic or random --
either with classical $m=3$ component spins, or else a 3-colouring: thus,
$\PP(\rr)$ has spin or flavour indices.~\footnote{
%%%%%%%%%%%%%%%%%%
The flavour case is the large-$S$ quantum model at temperatures low enough
for coplanarity, but high enough to ignore the small energies selecting
specific spin patterns~\cite{clh-pyrice} as in Sec.~\ref{sec:Heff}.}
%%%%%%%%%%%%%%%%%%
But I'll work out, instead, the simpler case of a dimer covering on the
depleted lattice; similar results are expected for the real problems.

Let the dimer pattern's polarization field be $\PP(\rr)$ 
on the depleted lattice ($z=3$), or $\EE(\rr)$ as a covering of 
the original ($z=4$) lattice; and let $\DD(\rr)$ be the
polarization field of the ($z=4$) dimers defining 
the depletion pattern.  It is trivial to check that 
   \be
           \PP = \frac{9}{8} (\EE+ \frac{1}{3} \DD).
   \label{eq:convertPE}
   \eeq
Now, the random depleted lattice is a ``quenched''
model in which first $\DD$ is fixed according to a single-component ensemble,
i.e. Eq.~\eqr{eq:entropy} with 
$\alpha |\PP|^2 \to \alphaD |\DD|^2$. 
%% and then $\EE(r)$ fluctuates
%% with $\DD$ acting as a kind of random field.
But consider, just for now, a random ensemble with two (equivalent) colours of 
non-overlapping dimers described by $\DD$ and $\EE$.  By cubic lattice
symmetry as well as $\DD \leftrightarrow \EE$ symmetry, 
the joint entropy density must be isotropic to quadratic order:
   \be
   \sigmaDE = \sigmaDE_0 - 
         \frac{1}{2} \alphaDE \Big(|\DD|^2 + |\EE|^2 + 2 \lambda \DD\cdot \EE\Big).
   \label{eq:entropy-joint}
   \eeq
where $0< \lambda <1$ is expected.
%%%%%%%%%%%%
\OMIT{(The argument for $\lambda >0$: it's harder
to avoid overlaps when both dimer coverings have the same polarization.
To rationalize $\lambda<1$, imagine a two-stage construction
starting from a loop ($z'=2$) covering with one colour
-- and a loop fugacity 2 -- in which case the polarization is proportional
to $\DD+\EE$; when we 2-colour each loop, the entropy conditioned on
$\DD+\EE$ is maximum when $\DD=\EE$.)}
%%%%%%%%%%%%
This also must be the entropy density for $\EE$ conditioned on a {\it frozen} 
(not untypical) $\DD(\rr)$.  
%%%%%%%%%%%%%%%%%
\OMIT{We get $\sigmaE=\sigmaE_0 - \frac{1}{2} \alpha \Big(\EE + \lambda \DD)^2$, 
with 
$\sigmaE_0 = \sigmaDE_0 - \frac{1}{2} \alphaDE (1-\lambda^2)|\DD|^2$.}
%%%%%%%%%%%%%%%%%
Using the change of variables \eqr{eq:convertPE}, 
%%%%%%%%%%%%%%%%%
\OMIT{(Namely, $\EE$ is $8/9 (\PP + 9/8(\lambda -1/3) \DD$.)}
%%%%%%%%%%%%%%%%%
the entropy density is \eqr{eq:entropy}
with $\PP\to \tilde{\PP} \equiv \PP + \tlambda \DD$
$\alpha\equiv \frac{8}{9} \alphaDE$ and
$\tlambda\equiv \frac{9}{8} (\lambda -\frac{1}{3})$.
Thus, $\PP$ is a linear combination of two independent fields $\tilde{\PP}(\rr)$ 
and $\DD(\rr)$, each having correlations like \eqr{eq:pseudodip}, with 
respective prefactors $1/\alpha$ and $1/\alphaD$.
On the other hand, a {\it periodic} depleted lattice
has a uniform $\DD(\rr)$, offsetting $\PP(\rr)$ by a constant.
Note that, unless a depleted lattice has cubic symmetry
or small $\DD$, the coefficients of $P_a P_b$ terms become anisotropic, 
(deriving from $O(DDEE)$ cross terms in \eqr{eq:entropy-joint}).
%%%%%%%%%%%%%%
\OMIT{It is no surprise to find long-range correlations even with maximal 
quenched disorder: the polarization is conserved at each vertex, 
just as in the parent pyrochlore lattice.}

\section{Conclusion}

I have shown there exist many depleted lattices (Sec.~\ref{sec:examples})
and given a picture
both of the ordered ground state (in Sec.~\ref{sec:ground})
and the cooperative paramagnet state (Sec.~\ref{sec:coop-para}).
In the latter state, the disconnected correlation function
of $\PP(\rr)$ has the pseudodipolar form \eqr{eq:pseudodip}.
Most of the results apply both to regular and random depleted lattices.
Random depletion is an inviting way to include just enough 
constraints to maintain many features of uniform lattices,
e.g. satisfying {\it every} triangle and allowing coplanar
ground states, which are violated in rare places in the case
of unconstrained site dilution~\cite{Hen01}.

\ack
I thank M. Lawler and C. Broholm for discussions.
This work was supported by NSF Grant No. DMR-0552461.

\end{document}